\documentclass{article}
\pdfoutput=1
\usepackage{spconf,amsmath,graphicx}

\usepackage[dvipsnames]{xcolor}

\newcommand*\linkcolours{ForestGreen}

\usepackage{times}
\usepackage{graphicx}
\usepackage{amssymb}
\usepackage{gensymb}
\usepackage{amsmath}
\usepackage{breakurl}

\usepackage{url,hyperref}
\hypersetup{
colorlinks,
linkcolor=\linkcolours,
citecolor=\linkcolours,
filecolor=\linkcolours,
urlcolor=\linkcolours}

\usepackage{algorithm}
\usepackage{algorithmic}

\usepackage[labelfont={bf},font=small]{caption}

\usepackage{mathtools, cuted}

\usepackage[noadjust, nobreak]{cite}

\usepackage{tabularx}
\usepackage{amsmath}

\usepackage{float}

\usepackage{pifont}

\newcommand*\GitHubLoc{https://github.com}

\newcolumntype{Y}{>{\centering\arraybackslash}X}

\usepackage[]{placeins}

\usepackage{placeins}

\usepackage{tikz}

\usepackage[framemethod=tikz]{mdframed}

\usepackage{afterpage}

\usepackage{stfloats}

\usepackage{atbegshi}
\newcommand{\handlethispage}{}
\newcommand{\discardpagesfromhere}{\let\handlethispage\AtBeginShipoutDiscard}
\newcommand{\keeppagesfromhere}{\let\handlethispage\relax}
\AtBeginShipout{\handlethispage}

\usepackage[colorinlistoftodos]{todonotes}

\usepackage{comment}

\title{EFFICIENT ATTENTION GUIDED 5G POWER AMPLIFIER DIGITAL PREDISTORTION}
\name{Alexandru Cioba$^{\star}$, Alvin Chua$^{\star}$, Da-shan Shiu$^{\star}$, Ting-Hsun Kuo$^{\dagger}$, Chia-Sheng Peng$^{\dagger}$}
\address{$^{\star}$ MediaTek Research \qquad $^{\dagger}$ MediaTek}
\begin{document}

%
\maketitle

\begin{abstract}

We investigate neural network (NN) assisted techniques for compensating the non-linear behaviour and the memory effect of a 5G PA through digital predistortion (DPD). Traditionally, the most prevalent compensation technique computes the compensation element using a Memory Polynomial Model (MPM). Various neural network proposals have been shown to improve on this performance. However, thus far they mostly come with prohibitive training or inference costs for real world implementations. In this paper, we propose a DPD architecture that builds upon the practical MPM formulation governed by neural attention. Our approach enables a set of MPM DPD components to individually learn to target different regions of the data space, combining their outputs for a superior overall compensation. Our method produces similar performance to that of higher capacity NN models with minimal complexity. Finally, we view our approach as a framework that can be extended to a wide variety of local compensator types.

\end{abstract}

\begin{keywords}
Predistortion, neural network, 5G PA
\end{keywords}
%

\section{Introduction}\label{intro}

Operating in an open-loop manner, RF power amplifiers (PAs) can naturally exhibit non-linear distortion. A common compensation concept, called digital predistortion (DPD), manipulates the input to the PA so that the actual output matches the desired output as much as possible. Several classical methods exist; notable among them is the memory polynomial model (MPM) family of functions. Lately, as the required operating bandwidth of PA becomes greater, e.g. in 5G radio, PAs can exhibit higher distortion effect. MPM solutions have thus been subjected to many extensions.



Spurred by the advances in Artificial Neural Network (NN) techniques, there has been increased interest in applying these methods for predistortion. Various promising architectures were proposed in \cite{5340581} to address the inverse problem. Due to its power and simplicity this architecture has remained relevant as a NN DPD benchmark, with variants being developed to model non-causal effects as in \cite{TADNN} and having been employed in recent studies such as \cite{NNMIMO}.\\
\indent In general, though compensation performance of NN assisted techniques can exceed those of conventional counterparts, their practicality has been limited due to the computational complexity of training NN models and furthermore, the prohibitive complexity of inference in the context of low power applications in mobile devices. One of the most successful methods involves modelling the PA itself by a NN and training a NN DPD inline with the pretrained PA model as in \cite{NNPANNDPD}, where this approach benefits from its own FPGA implementation.\\
\indent It has been claimed \cite{MIMO} that the open-loop DPD learning is responsible for introducing noise from the feed-back receiver into the DPD estimation, an undesirable effect. In this paper, we propose a neural-attention assisted DPD solution, which leverages the computational minimalism and generalization capability of the MPM together with the flexibility and noise robustness of NN models to address these issues. By using the highly efficient MPM as subcompensators, our method produces performance similar to high capacity NN  models, yet requires greatly reduced complexity. Furthermore, it inherits the performance  guarantees  of  simple MPMs. 
\section{Acknowledgements}
The authors wish to thank Alberto Bernacchia, Kader Medles and Abdellatif Salah for helpful comments and suggestions.

\section{Background}
\label{background}

\subsection{Notation}\label{notation}

\begin{figure}[h]
\centering
\includegraphics[width=0.7\columnwidth]{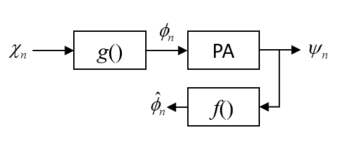}
\caption{Block diagram and notation for a DPD system.}
\label{DPD1}
\end{figure}

A DPD setup starts with an input complex-valued sequence $\chi_n$, representing the target digitally-modulated waveform. The signal flows through a digital predistorter (DPD) $g(\cdot)$ to produce another complex-valued sequence $\phi_n$. This second waveform is then passed through the analog PA, the output of which is a third signal $\psi_n$. An optimal DPD function $g^*(\cdot)$ should produce $\psi_n = C \chi_n$ up to some fixed scaling constant $C$, often referred to as the gain. As outlined in \cite{ILA}, a framework to solve for the DPD function, suitable in real time, is the inverse learning architecture (ILA). In ILA, one learns an (approximate) inverse transformation from the output of the analog PA, $\psi_n$, back to its input $\phi_n$. This transformation is denoted as $f(\cdot)$, and the output sequence of $f$ is denoted as $\hat{\phi}_n$. This inverse transform is subsequently used as the DPD function. The system and notation is given in Figure \ref{DPD1}. With a high quality learning algorithm, the concatenation of the ILA DPD function with the analog PA should produce $\psi_n \approx C \chi_n$ with little error. 

\subsection{Memory polynomial model}\label{sec:mpm}

As mentioned in Section \ref{intro}, MPM has seen widespread and continued use. Using MPM for DPD  was inspired by the Volterra series formulation of the underlying distortion phenomenon. Given, $\psi_n$, the MPM formulation uses the following polynomials as \emph{basis elements}: 
\begin{equation} \label{eq:basis}
\psi_{n-l} \cdot |\psi_{n-l}|^{2k},
\end{equation}
where $k = 0, 1, 2 ...$ is a variable degree of nonlinearity, and $l$ is a variable degree of delay. The MPM computes a weighted sum of basis elements to form an estimate of the PA input,  
\begin{equation} \label{eq:1}
\hat{\phi}_n = f_{\text{MPM}}(\psi_n) = \sum_{l,k} \lambda_{lk} \cdot \psi_{n-l} \cdot |\psi_{n-l}|^{2k},
\end{equation}
for some choice of complex coefficients $\lambda_{lk}$.




As indicated in Section \ref{notation}, the values of $\lambda_{lk}$ can be obtained by matching two signals $\phi$ and $\hat{\phi}$. This is a problem that can be phrased as a standard linear regression. Thus, any appropriate linear regression algorithms can be applied to solve for $\lambda_{lk}$. Particularly, if one wishes to minimize the expected squared distance between aligned sequences $\{ \phi_n \}$ and $\{ f_{\text{MPM}}(\psi_n) \}$, the least square (LS) solution applies.

The nonlinear basis elements used by the $f_{\text{MPM}}$ can be collected in a matrix $\mathbf{\Psi}$ with entries $\mathbf{\Psi}_{l,k}^n = \psi_{n-l} \cdot |\psi_{n-l}|^{2k}$, where $l$ denotes the delay dimension and and $k$ the nonlinear order dimension in (\ref{eq:1}). The target sequence $\{ \phi_n \}$ can be represented by a vector $\Phi$. The vector $\Lambda_{\text{LS}}$ that represents the coefficients $\{ \lambda_{lk}^n \}$ is solved by

\begin{equation} \label{eq:solving_lambda}
\Lambda_{\text{LS}} = (\mathbf{\Psi}^\dagger \mathbf{\Psi}
)^{-1}\mathbf{\Psi}^\dagger \Phi.
\end{equation}

\subsection{Performance metric}

As is the common practice in the DPD literature \cite{UnderstandingNMSE}, in this work, we evaluate the performance of various DPD methods by the normalized mean squared error (NMSE) between aligned PA input signal $\phi_n$ and the output of the DPD function, $\hat{\phi}_n = f(\psi_n)$. The definition for NMSE is as follows: 
\begin{equation}
\text{NMSE}(\hat{\phi}, \phi) = 10 \log_{10}(||\hat{\phi} - \phi||_2^2 / ||\phi||_2^2).
\end{equation}




\section{Neural attention assisted DPD}
\label{algo}


We present an architecture that intends to leverage the generalization capability of the MPM, while improving its predistortion capability and  flexibility. We refer to our architecture as an Attention Guided Memory Polynomial Neural Network (AGMPNN). Efforts have previously been made to combine the robustness and efficiency of the MPM with NN assisted solutions in different settings: typically trained in a closed-loop manner and revolved around modelling the MPM by a very small multilayer perceptron \cite{DLANN}, or cascading a NN DPD model together with the MPM to boost its predistortion capability. Our solution is adapted for the ILA and is an end-to-end DPD module.


\subsection{Amplitude-offset memory polynomial}

Typically, an MPM is used to cover the entire range of a signal. Here, we desire to have a number of MPM-like compensators that can be individually tuned for different regions of the input signal amplitude dimension. To achieve this effect, we augment an MPM in (\ref{eq:basis}) with an additional real-valued trainable amplitude offset parameter. We refer to such an augmented MPM as an amplitude-offset MPM (AOMPM).
The output of a single AOMPM with bias $b \in \mathbb{R}$ is,  
\begin{equation}\label{eq:AOMPM}
\hat{\phi}_n = \sum_{l,k} \lambda_{lk} \cdot \psi_{n-l} \cdot (|\psi_{n-l}| + b)^{2k}
\end{equation}



\subsection{Global compensation by assemblying local compensation}

Our approach follows the principle of the \emph{mixture of experts} framework \cite{Bishop:1995:NNP:525960}. The key concepts are:


\subsubsection{Local learning}
It is well known that for certain PA settings simple and explainable formulations like the MPM  achieve good results. While the MPM
promises to learn well within small amplitude-time bands, its performance degrades quickly if charged to cover a wide amplitude range. Capitalizing on this insight, we deploy several AOMPM learners, optimizing their compensation effect in individual amplitude-time regimes. 

\subsubsection{Attention}

\begin{figure}[h]
\centering
\includegraphics[width=0.8\columnwidth]{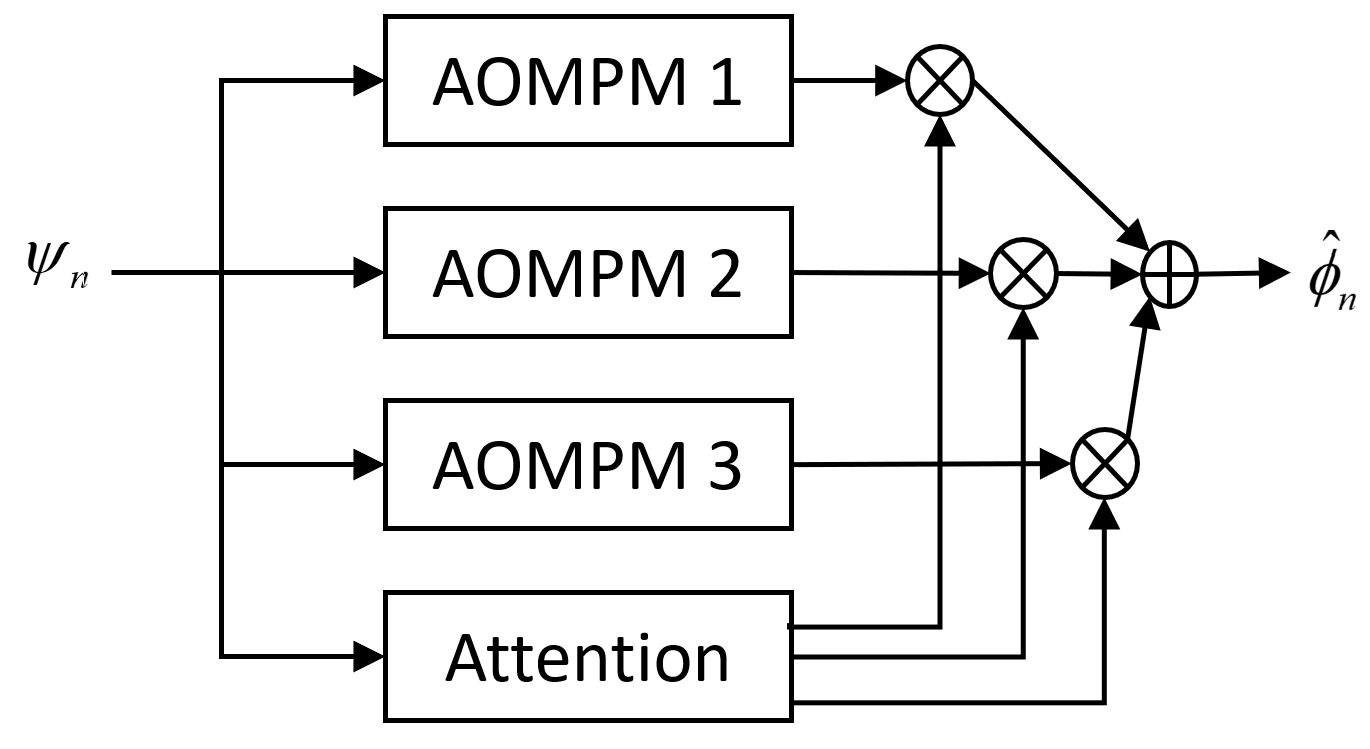}
\caption{Ensemble network architecture with three AOMPM.}
\label{smallnet}
\end{figure}

We use an attention mechanism to coordinate the aforementioned ensemble of MPM-like local learners. Through end-to-end training, the learners partition the entire amplitude-time domain into "subregions" to which they fit their parameters to independently. Additionally, the attention layer jointly learns with the local learners how best to summarize the outputs from these local learners into the final input to the PA. 
The full architecture is illustrated in Figure \ref{smallnet}. The architecture consists of $M$ AOMPMs, an attention head, and an output layer. The $m$-th AOMPM uses an offset $b^{(m)}$, $m = 1, ..., M$. From its observation of the amplitude of the input signal, the attention head dynamically assigns a weight to each AOMPM. Using such weights, the output layer forms a linear combination of the outputs of the AOMPM as the final output. The attention head produces the weights $w^{(m)}$ for AOMPM $m$ with the following rule:
\begin{align}
\operatorname{ReLU^\dagger}(a;b) &= \max(0, |a| + b)\\
\operatorname{MLP}_m(\psi) &=  \sum_l  \mu_l^{(m)} \cdot \operatorname{ReLU^\dagger}(\psi_{n-l};b^{(m)}) + \nu_l^{(m)}\\
w^{(m)} &=  \operatorname{softmax}_m \left( \operatorname{MLP}_m(\psi) \right)
\end{align}
where $b^{(m)}$ is as in (\ref{eq:AOMPM}), and $\mu_l^{(m)}$ and $\nu_l^{(m)}$ are trainable real-valued parameters, operating on the $l$-th tap of signal $\phi_n$. We note that, while the values of $b^{(m)}$, $\mu_l^{(m)}$, and $\nu_l^{(m)}$ are learned during a training phase, the values of $w_n^{(m)}$ vary dynamically in time during inference time. Finally, the output layer produces the overall predistorter output as 
\begin{equation}\label{eq:final}
\hat{\phi}_n = \sum_{m} w^{(m)} \left( \sum_{l,k} \lambda^{(m)}_{lk} \cdot \psi_{n-l} \cdot (|\psi_{n-l}| + b^{(m)})^{2k} \right)
\end{equation}

\subsection{Model complexity}

The complexity of our model depends on the number of taps, $L$, the maximum order on squared amplitude, $K$, and the number of AOMPM employed, $M$. Our model comprises of $4LKM + LM + 4L + 2M + 2$ trainable real parameters. In our model, the number of trainable parameters is almost identical to the number of multiply-and-accumulate operations, making  the number of parameters a good metric to represent real-time inference cost.
In this work, we use a very simple MLP to compute the attention coefficients $w_n^{(m)}$. We plan to investigate further attention variants in future work.


\subsection{Determination of model parameters}

The AOMPM parameters $\lambda_{l,k}^{(m)}$ and $b^{(m)}$, and attention paramters $\mu_l^{(m)}$ and $\nu_l^{(m)}$ can be learned online. The entire system in Figure \ref{smallnet} can be trained end-to-end to minimize the MSE between $\phi$ and $\hat{\phi}$, using typical optimizers for NNs.

\subsection{Performance lower bound}


A common factor that hinders the adoption of NN-based solutions is that the behaviour and generalizability of neural networks tends to be hard to explain. When performance drops unexpectedly, it can be difficult to formulate a solution quickly. 
In contrast, the MPM holds credibility from a long history of industrial use. 
Our model is built on top of an ensemble of AOMPMs. As such it is guaranteed to perform at least as well as a standard MPM. Furthermore, 
the additional risk only lies in the simple attention layer, a property which renders our solution light-weight and interpretable.

\begin{figure*}[tbh]
\centering
\includegraphics[width=0.9\textwidth]{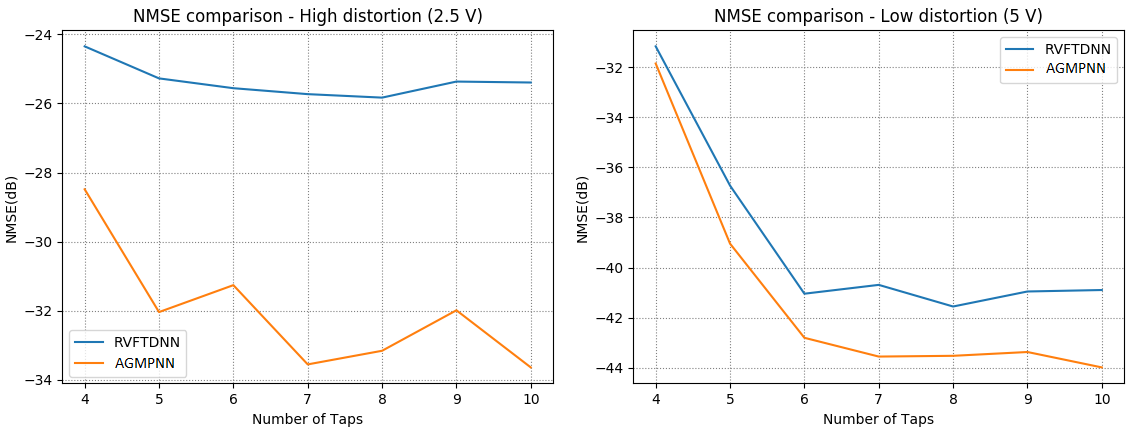}
\caption{DPD Performance as a function of maximum number of input delay taps.}
\label{vs_taps}
\end{figure*}

\section{Experiments}
\label{experiments}




\begin{figure*}[tbh]
\centering
\includegraphics[width=0.9\textwidth]{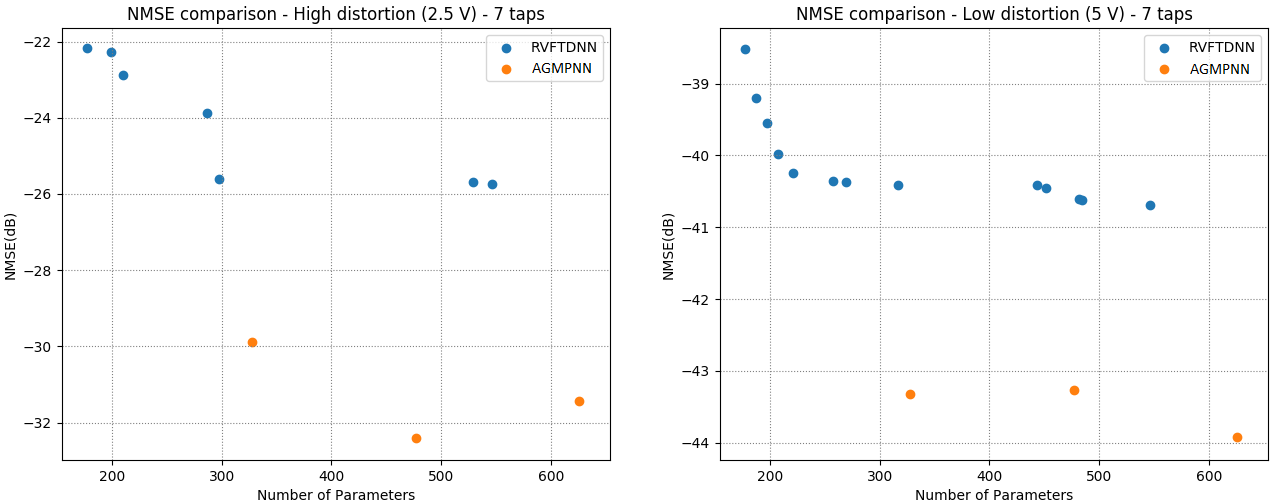}
\caption{DPD performance in NMSE at low and high distortion regimes. All DPD schemes have 7 memory taps at their inputs.}
\label{vs_complexity}
\end{figure*} 

\subsection{Methodology}

In this paper, we evaluate various DPD schemes using a 5G PA that exhibits a non-trivial memory effect.
\begin{table}[htbp]
  \centering
  \caption{Optimal architecture parameters for RVFTDNN, vs. input taps, optimized for two distortion levels. $N_1$ and $N_2$ represent the number of neurons in the first and second hidden layers, respectively. }
    \begin{tabular}{ cc }
    \textbf{High distortion} & \textbf{Low distortion} \\
        \begin{tabular}{| c | c | c | c |}
        
        \hline
        \textbf{Taps} & \textbf{$N_1$} & \textbf{$N_2$} \\ \hline
        4     & 17 & 15 \\ \hline
        5     & 13 & 13 \\ \hline
        6     & 18 & 17 \\ \hline
        7     & 16 & 16 \\ \hline
        8     & 19 & 12  \\ \hline
        9     & 13 & 17 \\ \hline
        10    & 16 & 11 \\ \hline
        \end{tabular} &
        
        \quad

        \begin{tabular}{| c | c | c | c |}
        \hline
        \textbf{Taps} & \textbf{$N_1$} & \textbf{$N_2$}\\ \hline
        4     & 17 & 17 \\ \hline
        5     & 18 & 18 \\ \hline
        6     & 15 & 10 \\ \hline
        7     & 16 & 16 \\ \hline
        8     & 15 & 10  \\ \hline
        9     & 16 & 12 \\ \hline
        10    & 15 & 14 \\ \hline
        \end{tabular} \\
    \end{tabular}
    \label{tab:baselines5}%
\end{table}


In the following experiments, we use the real-valued focused time-delay neural network (RVFTDNN) of \cite{5340581} as a baseline. To compare various DPD schemes fairly, in our evaluations we carefully control the computational complexity at inference time (and the number of non-causal taps of the input waveform to a DPD method). For all considered model families, we perform an exhaustive architecture search across all network architecture parameters. We limit the total number of trainable parameters to be within 100 to 600. Table \ref{tab:baselines5} presents the architecture parameters that result in the best NMSE at the low-distortion and high-distortion regimes. In subsequent experiments, for any given operating point, we employ its corresponding  optimal architecture. 



\label{training}


For our experiment, we capture data from a 5G PA at both the low and high distortion levels. Captured signals are split into contiguous segments of discrete samples. We align a discrete sample segment with its ground truth target. Training is performed with a batch size $50$ of such captured segment-ground truth pairs. The well known \emph{Adam} \cite{kingma2014adam} optimizer is used to minimize the NMSE. An early stopping criterion was used, which resulted in training epochs in between $20$ and $30$ on average. We observed no overfitting to the training set.



\subsection{Results}





Figure \ref{vs_complexity} presents the performance-complexity trade-off. For all DPD schemes, we allow 7 input memory taps. One can see clearly from the results that, given the same complexity,  
our model achieves a much lower NMSE than the baseline at both the low distortion and high distortion regimes. Not shown in Figure \ref{vs_complexity} is that, even though both models have the same number of parameters, our model achieves such a performance after a comparatively shorter training time. 
Figure \ref{vs_taps} compares the performance of the DPD methods over the total number of input delay taps, $L+1$. Here, for each value of $L$, the RVFTDNN is optimized within a capacity of 600 parameters. The AGMPNN comprises $43L+51$ parameters.
We observe that for all values of $L$, our method consistently outperforms RVFTDNN, by as much as 3.5 dB.

\section{Conclusions}
\label{conclusions}

We investigate neural network (NN) assisted techniques for compensating the non-linear behaviour and the memory effect of a 5G PA through digital predistortion (DPD). We presented a DPD architecture that builds upon the practical MPM governed by neural attention. Our approach enables a set of AOMPM components to individually learn to target different scenarios, and to collectively learn to combine their outputs for an ideal overall compensation. Our results indicate that AGMPNN is able to achieve far better predistortion performance in terms of NMSE compared to densely connected neural networks such as RVFTDNN, and does so consistently across multiple PA settings, proving increased robustness.




\bibliographystyle{IEEEbib}
\bibliography{bibliography}

\begin{thebibliography}{10}

\bibitem{5340581}
M.~{Rawat}, K.~{Rawat}, and F.~M. {Ghannouchi},
\newblock ``Adaptive digital predistortion of wireless power
  amplifiers/transmitters using dynamic real-valued focused time-delay line
  neural networks,''
\newblock {\em IEEE Transactions on Microwave Theory and Techniques}, vol. 58,
  no. 1, pp. 95--104, Jan 2010.

\bibitem{TADNN}
T.~{Gotthans}, G.~{Baudoin}, and A.~{Mbaye},
\newblock ``Digital predistortion with advance/delay neural network and
  comparison with volterra derived models,''
\newblock in {\em 2014 IEEE 25th Annual International Symposium on Personal,
  Indoor, and Mobile Radio Communication (PIMRC)}, Sep. 2014, pp. 811--815.

\bibitem{NNMIMO}
P.~{Jaraut}, M.~{Rawat}, and F.~M. {Ghannouchi},
\newblock ``Composite neural network digital predistortion model for joint
  mitigation of crosstalk,$i/q$imbalance, nonlinearity in mimo transmitters,''
\newblock {\em IEEE Transactions on Microwave Theory and Techniques}, vol. 66,
  no. 11, pp. 5011--5020, Nov 2018.

\bibitem{NNPANNDPD}
Chance Tarver, Alexios Balatsoukas-Stimming, and Joseph~R. Cavallaro,
\newblock ``Design and implementation of a neural network based predistorter
  for enhanced mobile broadband,''
\newblock {\em arXiv preprint arXiv:1907.00766}, 2019.

\bibitem{MIMO}
M.~{Abdelaziz}, L.~{Anttila}, and M.~{Valkama},
\newblock ``Reduced-complexity digital predistortion for massive mimo,''
\newblock in {\em 2017 IEEE International Conference on Acoustics, Speech and
  Signal Processing (ICASSP)}, March 2017, pp. 6478--6482.

\bibitem{ILA}
{Changsoo Eun} and E.~J. {Powers},
\newblock ``A new volterra predistorter based on the indirect learning
  architecture,''
\newblock {\em IEEE Transactions on Signal Processing}, vol. 45, no. 1, pp.
  223--227, Jan 1997.

\bibitem{UnderstandingNMSE}
P.~{H{\"a}ndel},
\newblock ``Understanding normalized mean squared error in power amplifier
  linearization,''
\newblock {\em IEEE Microwave and Wireless Components Letters}, vol. 28, no.
  11, pp. 1047--1049, Nov 2018.

\bibitem{DLANN}
X.~{Feng}, B.~{Feuvrie}, A.~S. {Descamps}, and Y.~{Wang},
\newblock ``Digital predistortion method combining memory polynomial and
  feed-forward neural network,''
\newblock {\em Electronics Letters}, vol. 51, no. 12, pp. 943--945, 2015.

\bibitem{Bishop:1995:NNP:525960}
Christopher~M. Bishop,
\newblock {\em Neural Networks for Pattern Recognition},
\newblock Oxford University Press, Inc., New York, NY, USA, 1995.

\bibitem{kingma2014adam}
Diederik~P Kingma and Jimmy Ba,
\newblock ``Adam: A method for stochastic optimization,''
\newblock {\em arXiv preprint arXiv:1412.6980}, 2014.

\end{thebibliography}

\clearpage

\end{document}